\documentclass[12pt]{article}
\usepackage{amsmath}
\usepackage[pdftex,unicode]{hyperref}
\usepackage{cmap}
\def\eqs#1{\begin{equation}\begin{split}#1\end{split}\end{equation}}

\def\baselinestretch{1.5}
\numberwithin{equation}{section}
\begin{document}

\begin{center}
{\Large Affine Lie algebras, Lax pairs  and integrable discrete and continuous  systems}
\end{center}

\begin{center}
{Rustem  Garifullin}\footnote{e-mail:  grustem@gmail.com} and {Ismagil Habibullin}\footnote{e-mail: habibullinismagil@gmail.com}\\

{Ufa Institute of Mathematics, Russian Academy of Science,\\
Chernyshevskii Str., 112, Ufa, 450077, Russia}\\

\end{center}

\begin{abstract}
A consistent set of six integrable discrete and continuous dynamical systems are suggested corresponding to arbitrary affine Lie algebra. The set contains a system of partial differential equations which can be treated as a version of generalized Toda lattice while  semi-discrete systems in the set define the Backlund transform for this Toda lattice and the fully discrete representative of the set can be  obtained as a superposition of such kind Backlund transforms. Four linear Zakharov-Shabat type systems taken pairwise realize Lax pairs for these six dynamical systems.
\end{abstract}

\def\baselinestretch{1.5}


%\newpage
%\tableofcontents

\section{Introduction}

Lax representation is known to be one of the fundamental attributes of integrability. All the other properties of integrable equations like symmetries, conserved quantities, exact and asymptotic solutions can be algorithmically derived from the Lax operators. In the survey \cite{DrS} a deep connection between concept of Lax representation and affine Lie algebras was observed. It was shown that one can assign a generalized two-dimensional Toda lattice
to any Kac-Moody  type Lie algebra. Remarkably the related Zakharov-Shabat operators are completely described in algebraic terms. Moreover, even the formal quasi-classical approximation of eigenfunctions is constructed in a purely algebraic way. 

The purpose of the present article is to define discrete versions of Zakharov-Shabat eigenvalue problem related to affine Lie algebras and find the discrete and semi-discrete dynamical systems associated with these operators.

\section{General scheme}

Recall the necessary definitions. Let $G$ be a Kac-Moody algebra and the elements $\{e_i,f_i,h_i \}_{i=0}^{N}$ constitute a system of 
the canonical generators:
\begin{enumerate}
	\item $e_i\neq 0$, $f_i\neq 0$, $h_i\neq 0$ for any $i$;
	\item for any $i,j$ the relations hold
	$$[h_i,h_j]=0,$$
	$$[e_i,f_j]=\delta_{ij}h_i,$$
	$$[h_i,e_j]=N_{ij}e_j,$$
	$$[h_i,f_j]=-N_{ij}f_j.$$
\end{enumerate}
Here $N=\{N_{ij}\}$ is the Cartan matrix of the algebra, and $\delta_{ij}$ is the Kroneker symbol. Below we use the usual matrix representation of the algebra $G$, see \cite{DrS}.

Introduce the following four linear equations:
\begin{eqnarray}
P_{n+1,m}&=&U_{n,m}P_{n,m},\label{U}\\
P_{n,m+1}&=&V_{n,m}P_{n,m},\label{V}\\
\frac{d}{dx}P_{n,m}&=&X_{n,m}P_{n,m},\label{X}\\
\frac{d}{dy}P_{n,m}&=&Y_{n,m}P_{n,m},\label{Y}
\end{eqnarray}
where the potentials are of the form
\begin{eqnarray}
U_{n,m}&=&-\Lambda+\displaystyle{e^{F_{n+1,m}-F_{n,m}}},\label{UU}\\
V_{n,m}&=&E+\displaystyle{e^{F_{n,m+1}}\bar \Lambda e^{-F_{n,m}}},\label{VV}\\
X_{n,m}&=&-\displaystyle{e^{F_{n,m}}\bar \Lambda e^{-F_{n,m}}},\label{XX}\\
Y_{n,m}&=&-\Lambda+\displaystyle{\frac{d}{dy}F_{n,m}}.\label{YY}
\end{eqnarray}
Operators (\ref{XX})-(\ref{YY}) are similar to those studied in \cite{DrS} and operators (\ref{UU})-(\ref{VV}) generalize difference operators introduced in  the case one independent variable \cite{Suris}. 
Explain the notations in (\ref{UU}-\ref{YY}). Here $E$ is the identity matrix, $\Lambda=\sum_{i=0}^{N} e_i$ and $\bar\Lambda=\sum_{i=0}^{N} f_i$. In order to describe the functional parameter $F=F_{n,m}$ introduce the linear space $V_d$ of all diagonal matrices of the corresponding order.  Then set $V_0=\Lambda V_d\bar\Lambda$. It can be proved that $V_0$ is a commutative ring satisfying the condition $\Lambda V_0\bar\Lambda \subset V_0$ as well as the condition $\bar\Lambda V_0\Lambda \subset V_0$. We suppose that $F=F_{n,m}$ is an arbitrary element of the space $V_0$. Compatibility conditions of the equations (\ref{U}-\ref{Y}) taken pairwise generate six nonlinear equations:
\begin{eqnarray}
e^{F_{n+1,m+1}-F_{n,m+1}}-e^{F_{n+1,m}-F_{n,m}}&=&\Lambda e^{F_{n,m+1}}\bar \Lambda e^{-F_{n,m}}- e^{F_{n+1,m+1}}\bar \Lambda e^{-F_{n+1,m}}\Lambda,\label{UV}\\
F_{n,m,x,y}&=&e^{F_{n,m}}\bar \Lambda e^{-F_{n,m}}\Lambda-\Lambda e^{F_{n,m}}\bar \Lambda e^{-F_{n,m}},\label{XY}\\
(F_{n+1,m,x}-F_{n,m,x})e^{F_{n+1,m}-F_{n,m}}&=&e^{F_{n+1,m}}\bar \Lambda e^{-F_{n+1,m}}\Lambda-\Lambda e^{F_{n,m}}\bar \Lambda e^{-F_{n,m}},\label{XU}\\
F_{n,m+1,y}-F_{n,m,y}&=&\Lambda e^{F_{n,m+1}}\bar \Lambda e^{-F_{n,m}}-e^{F_{n,m+1}}\bar \Lambda e^{-F_{n,m}}\Lambda,\label{YV}\\
(F_{n,m+1,x}-e^{F_{n,m}-F_{n,m+1}})\bar \Lambda&=&\bar \Lambda(F_{n,m,x}-e^{F_{n,m}-F_{n,m+1}}),\label{XV}
\\
(F_{n+1,m,y}-e^{F_{n,m}-F_{n+1,m}})\Lambda&=&\Lambda(F_{n,m,y}-e^{F_{n,m}-F_{n+1,m}}).\label{YU}
\end{eqnarray}
Here $F_{n,m,x}:=\frac{\partial}{\partial x}F_{n,m}$, $F_{n,m,y}:=\frac{\partial}{\partial y}F_{n,m}$ and $F_{n,m,x,y}:=\frac{\partial^2}{\partial x\partial y}F_{n,m}$. It can be proved that all the equations (\ref{UV})-(\ref{YU}) are self consistent, they really define dynamical systems: first one is fully discrete, the second is fully continuous and the others are semi-discrete. The following statement is a straightforward consequence of the construction method 
of equations (\ref{UV})-(\ref{YU}):

{\bf Theorem.} Operators $D_n$, $D_m$, $D_x$, $D_y$ mutually commute on the set of the dynamical variables $F_{n,m}$, $F_{n,m,x}$, $F_{n,m,y}$. Where $D_x$ and  $D_y$ are operators of total differentiation with respect to $x$ and $y$, operators $D_n$, $D_m$ act due to the rule $D_nF_{n,m}=F_{n+1,m}$, $D_mF_{n,m}=F_{n,m+1}$.

{\bf Corollary of Theorem.} Pair of semi-discrete equations (\ref{XU}), (\ref{YU}) as well as pair of equations (\ref{YV}), (\ref{XV}) define a sequence of the Backlund transformations  for equation (\ref{XY}). Equation (\ref{UV}) follows from the superposition principle for these Backlund transforms.

Equation (\ref{XY}) admits a reduction, defined by the formula  $F=\sum_{i=0}^{N} u_ih_i$ which leads to the usual generalized Toda lattice, corresponding algebra $G$ (see, \cite{DrS}). Unfortunately as a rule this reduction is not compatible with the other equations from the list (\ref{UV})-(\ref{YU}). This is one  of the reasons why the discretization problem for the Drinfel'd-Sokolov formalism is still unsolved. Thus we come up to an important problem of describing reductions of equation (\ref{XY}) compatible simultaneously with all of the equations (\ref{UV})-(\ref{YU}).

\section{Examples of consistent sets of the dynamical systems}

In this section we give examples of the set of dynamical systems corresponding the algebras $A_1^{(1)}$, $D_3^{(2)}$, $B_2^{(1)}$.

\subsection{Dynamical systems corresponding $A_1^{(1)}$}

For $A_1^{(1)}$ the  matrices $\Lambda$ and $\bar \Lambda$ look as follows:
\eqs{\label{LA11}\Lambda=\lambda\left(\begin{array}{cc}0&1\\1&0\end{array}\right),\quad             \bar \Lambda=\frac{1}{\lambda}\left(\begin{array}{cc}0&1\\1&0\end{array}\right),}
where $\lambda$ is interpreted as a spectral parameter, $F_{n,m}$ is a diagonal matrix:
\eqs{\label{FA11}F_{n,m}=\left(\begin{array}{cc}u^1_{n,m}&0\\0&u^2_{n,m}\end{array}\right).}
Equations \eqref{UV}-\eqref{YU} take the form:
\eqs{\label{UVA11}
e^{u^1_{n+1,m+1}-u^1_{n+1,m}-u^1_{n,m+1}+u^1_{n,m}}=\frac{1+e^{u^2_{n,m+1}-u^1_{n+1,m}}}{1+e^{u^1_{n,m+1}-u^2_{n+1,m}}},\\
e^{u^2_{n+1,m+1}-u^2_{n+1,m}-u^2_{n,m+1}+u^2_{n,m}}=\frac{1+e^{u^1_{n,m+1}-u^2_{n+1,m}}}{1+e^{u^2_{n,m+1}-u^1_{n+1,m}}}}
\eqs{\label{XYA11}
u^1_{n,m,x,y}=e^{u^1_{n,m}-u^2_{n,m}}-e^{u^2_{n,m}-u^1_{n,m}},\\
u^1_{n,m,x,y}=e^{u^2_{n,m}-u^1_{n,m}}-e^{u^1_{n,m}-u^2_{n,m}}}
\eqs{\label{UXA11}
u^1_{n+1,m,x}-u^1_{n,m,x}=e^{u^1_{n,m}-u^2_{n+1,m}}-e^{u^2_{n,m}-u^1_{n+1,m}},\\
u^2_{n+1,m,x}-u^2_{n,m,x}=e^{u^2_{n,m}-u^1_{n+1,m}}-e^{u^1_{n,m}-u^2_{n+1,m}}}
\eqs{\label{VYA11}
u^1_{n,m+1,y}-u^1_{n,m,y}=e^{u^2_{n,m+1}-u^1_{n,m}}-e^{u^1_{n,m+1}-u^2_{n,m}},\\
u^2_{n,m+1,y}-u^2_{n,m,y}=e^{u^1_{n,m+1}-u^2_{n,m}}-e^{u^2_{n,m+1}-u^1_{n,m}}}
\eqs{\label{VXA11}
u^1_{n,m+1,x}-u^2_{n,m,x}=e^{u^1_{n,m}-u^1_{n,m+1}}-e^{u^2_{n,m}-u^2_{n,m+1}},\\
u^2_{n,m+1,x}-u^1_{n,m,x}=e^{u^2_{n,m}-u^2_{n,m+1}}-e^{u^1_{n,m}-u^1_{n,m+1}}}
\eqs{\label{UYA11}
u^1_{n+1,m,y}-u^2_{n,m,y}=e^{u^1_{n+1,m}-u^1_{n,m}}-e^{u^2_{n+1,m}-u^2_{n,m}},\\
u^2_{n+1,m,y}-u^1_{n,m,y}=e^{u^2_{n+1,m}-u^2_{n,m}}-e^{u^1_{n+1,m}-u^1_{n,m}}}
All of the equations (\ref{UVA11})-(\ref{UYA11}) are compatible with the additional constraint $u^1_{n,m}+u^2_{n,m}=\alpha n+ \beta m$ which reduces the set of equations to a set of equations with the only unknown $u^1_{n,m}$. Here $\alpha$ and $\beta$ are constant parameters.

\subsection{Dynamical systems corresponding $D_3^{(2)}$}

In this case the matrices $\Lambda$ and $\bar \Lambda$ look as follows:
\eqs{\label{LD32}\Lambda=\lambda\left(\begin{array}{cccccc}0&0&0&1&0&0\\1&0&0&0&0&0\\0&1&0&0&0&0\\0&0&0&0&0&1\\0&0&1&0&0&0\\0&0&0&0&1&0\end{array}\right),\quad             \bar \Lambda=\frac{1}{\lambda}\left(\begin{array}{cccccc}0&1&0&0&0&0\\0&0&2&0&0&0\\0&0&0&0&2&0\\2&0&0&0&0&0\\0&0&0&0&0&1\\0&0&0&2&0&0\end{array}\right),}
where $\lambda$ is the spectral parameter, $F_{n,m}$ is at the very beginning a diagonal matrix with six functional parameters $F_{n,m}=diag(u^1_{n,m},u^2_{n,m},u^3_{n,m},u^4_{n,m},u^5_{n,m},u^6_{n,m})$. Then one can reduce the system of the form (\ref{UV})-(\ref{YU}) obtained to a set of three-field systems with:
\eqs{\label{FD32}F_{n,m}=\left(\begin{array}{cccccc}u^1_{n,m}&0&0&0&0&0\\0&u^2_{n,m}&0&0&0&0\\0&0&u^3_{n,m}&0&0&0\\0&0&0&u^3_{n,m}&0&0\\0&0&0&0&u^1_{n,m}&0\\0&0&0&0&0&u^2_{n,m}\end{array}\right).}
Thus the set of reduced equations \eqref{UV}-\eqref{YU} can be written in the following form:
\eqs{\label{UVD32}
{{\rm e}^{u^1_{{n+1,m}}-u^1_{{n,m}}}}-{{\rm e}^{u^1_{{n+1,m+1}}-u^2_{{n+1,m}}}}-{{\rm e}^{u^1_{{n+1,m+1}}-u^1_{{n,m+1}}}}+2\,{{\rm e}^{u^3_{{n,m+1}}-u^1_{{n,m}}}}=0\\
{{\rm e}^{u^2_{{n+1,m}}-u^2_{{n,m}}}}-2\,{{\rm e}^{u^2_{{n+1,m+1}}-u^3_{{n+1,m}}}}+{{\rm e}^{u^1_{{n,m+1}}-u^2_{{n,m}}}}-{{\rm e}^{u^2_{{n+1,m+1}}-u^2_{{n,m+1}}}}=0\\
{{\rm e}^{u^3_{{n+1,m}}-u^3_{{n,m}}}}-2\,{{\rm e}^{u^3_{{n+1,m+1}}-u^1_{{n+1,m}}}}+2\,{{\rm e}^{u^2_{{n,m+1}}-u^3_{{n,m}}}}-{{\rm e}^{u^3_{{n+1,m+1}}-
u^3_{{n,m+1}}}}=0
}
\eqs{\label{XYD32}
u^1_{{n,m},x,y}={{\rm e}^{u^1_{{n,m}}-u^2_{{n,m}}}}-2{{\rm e}^{-u^1_{{n,m}}+u^3_{{n,m}}}},\\
u^2_{{n,m},x,y}=2{{\rm e}^{-u^3_{{n,m}}+u^2_{{n,m}}}}-{{\rm e}^{u^1_{{n,m}}-u^2_{{n,m}}}},\\
u^3_{{n,m},x,y}=2{{\rm e}^{-u^1_{{n,m}}+u^3_{{n,m}}}}-2{{\rm e}^{-u^3_{{n,m}}+u^2_{{n,m}}}}}
\eqs{\label{UXD32}
u^1_{{n+1,m},x}-u^1_{{n,m},x}=-2{{\rm e}^{-u^1_{{n+1,m}}+u^3_{{n,m}}}}+{{\rm e}^{u^1_{{n,m}}
-u^2_{{n+1,m}}}},\\
u^2_{{n+1,m},x}-u^2_{{n,m},x}=-{{\rm e}^{u^1_{{n,m}}-u^2_{{n+1,m}}}}+2{{\rm e}^{u^2_{{n,m}}-u_{{3,n+1,m}}}},\\
u^3_{{n+1,m},x}-u^3_{{n,m},x}=-2{{\rm e}^{u^2_{{n,m}}-u^3_{{n+1,m}}}}+2{{\rm e}^{-u^1_{{n+1,m}}+u^3_{{n,m}}}}}
\eqs{\label{VYD32}
u^1_{{n,m+1},y}-u^1_{{n,m},y}={{\rm e}^{u^1_{{n,m+1}}-u^2_{{n,m}}}}+2{{\rm e}^{u^3_{{n,m+1}}-u^1_{{n,m}}}},\\
u^2_{{n,m+1},y}-u^2_{{n,m},y}=-2{{\rm e}^{u^2_{{n,m+1}}-u^3_{{n,m}}}}+{{\rm e}^{u^1_{{n,m+1}}-u^2_{{n,m}}}},\\
u^3_{{n,m+1},y}-u^3_{{n,m},y}=-2{{\rm e}^{u^3_{{n,m+1}}-u^1_{{n,m}}}}+2{{\rm e}^{u^2_{{n,m+1}}-u_{{3,n,m}}}}
}
\eqs{\label{VXD32}
u^1_{{n,m+1},x}-u^2_{{n,m},x}={{\rm e}^{-u^1_{{n,m+1}}+u^1_{{n,m}}}}-{{\rm e}^{u^2_{{n,m}}-
u^2_{{n,m+1}}}},\\
u^2_{{n,m+1},x}-u^3_{{n,m},x}={{\rm e}^{u^2_{{n,m}}-u_{{2,n,m+1}}}}-{{\rm e}^{u^3_{{n,m}}-u^3_{{n,m+1}}}},\\
u^3_{{n,m+1},x}-u^1_{{n,m},x}={{\rm e}^{u^3_{{n,m}}-u^3_{{n,m+1}}x,y}}-{{\rm e}^{-u^1_{{n,m+1}}+u^1_{{n,m}}}}}
\eqs{\label{UYD32}
u^1_{{n+1,m},y}-u^3_{{n,m},y}={{\rm e}^{u^1_{{n+1,m}}x,y-u^1_{{n,m}}}}-{{\rm e}^{u^3_{{n+1,m}}-u^3_{{n,m}}}},\\
u^2_{{n+1,m},y}-u^1_{{n,m},y}={{\rm e}^{u^2_{{n+1,m}}-u^2_{{n,m}}}}-{{\rm e}^{u^1_{{n+1,m}}-u^1_{{n,m}}}},\\
u^3_{{n+1,m},y}-u^2_{{n,m},y}={{\rm e}^{u^3_{{n+1,m}}-u^3_{{n,m}}}}-{{\rm e}^{u^2_{{n+1,m}}-u^2_{{n,m}}}}
}
It is remarkable that the system admits one more reduction. It is given by the constraint $u^1_{n,m}+u^2_{n,m}+u^3_{n,m}=\alpha n+ \beta m$, where $\alpha$ and $\beta$ are constants.

\subsection{Dynamical systems corresponding $B_2^{(1)}$}

For $B_2^{(1)}$ the matrices $\Lambda$ and $\bar \Lambda$ look as follows:
\eqs{\label{LB21}\Lambda=\lambda\left(\begin{array}{ccccc}0&0&0&1/2&0\\1&0&0&0&1/2\\0&1&0&0&0\\0&0&1&0&0\\0&0&0&1&0\end{array}\right),\quad             \bar \Lambda=\frac{1}{\lambda}\left(\begin{array}{ccccc}0&1&0&0&0\\0&0&2&0&0\\0&0&0&2&0\\2&0&0&0&1\\0&2&0&0&0\end{array}\right),}
Here the functional parameter is given by a non-diagonal matrix  $F_{n,m}$:
\eqs{\label{FB21}F_{n,m}=\left(\begin{array}{ccccc}u^1_{n,m}/2&0&0&0&u^1_{n,m}/4\\0&u^2_{n,m}&0&0&0\\0&0&u^3_{n,m}&0&0\\0&0&0&u^4_{n,m}&0\\u^1_{n,m}&0&0&0&u^1_{n,m}/2\end{array}\right).}
The matter is that the Dynkin diagram for $B_N^{(1)}$ contains  a vertex which is connected with three other vertices. That is why in this case space $V_0$ contains non-diagonal matrices and therefore $F$ is not diagonal. Me choose the eigenvalues of $F$ as field variables. Such a parametrization of $F$ is more convenient. 
Equations \eqref{UV}-\eqref{YU} take the form:
\eqs{\label{UVB21}
-2{{\rm e}^{u^1_{{n,m+1}}-u^2_{{n,m}}}}+{{\rm e}^{u^2_{{n+1,m+1}}-u^2_{{n,m+1}}}}-{{\rm e}^{u^2_{{n+1,m}}-u^2_{{n,m}}}}+2{{\rm e}^{u^2_{{n+1,m+1}}
-u^3_{{n+1,m}}}}=0,\\
{{\rm e}^{u^1_{{n+1,m+1}}-u^1_{{n,m+1}}}}-2{{\rm e}^{u^4_{{n,m+1}}-u^1_{{n,m}}}}+2{{\rm e}^{u^1_{{n+1,m+1}}-u^2_{{n+1,m}}}}-{{\rm e}^{u^1_{{n+1,m}}
-u^1_{{n,m}}}}=0,\\
-2{{\rm e}^{u^2_{{n,m+1}}-u^3_{{n,m}}}}+{{\rm e}^{u^3_{{n+1,m+1}}-u^3_{{n,m+1}}}}-{{\rm e}^{u^3_{{n+1,m}}-u^3_{{n,m}}}}+2{{\rm e}^{u^3_{{n+1,m+1}}-
u^4_{{n+1,m}}}}=0,\\
-2{{\rm e}^{u^3_{{n,m+1}}-u^4_{{n,m}}}}+{{\rm e}^{u^4_{{n+1,m+1}}-u^4_{{n,m+1}}}}+2{{\rm e}^{u^4_{{n+1,m+1}}-u^1_{{n+1,m}}}}-{{\rm e}^{u^4_{{n+1,m}}-u^4_{{n,m}}}}=0}
\eqs{\label{XYB21}
u^1_{{n,m},x,y}=2{{\rm e}^{u^1_{{n,m}}-u^2_{{n,m}}}}-2{{\rm e}^{-u^1_{{n,m}}+u^4_{{n,m}}}},\\
u^2_{{n,m},x,y}=2{{\rm e}^{-u^3_{{n,m}}+u^2_{{n,m}}}}-2{{\rm e}^{u^1_{{n,m}}-u^2_{{n,m}}}},\\
u^3_{{n,m},x,y}=2{{\rm e}^{u^3_{{n,m}}-u^4_{{n,m}}}}-2{{\rm e}^{-u^3_{{n,m}}+u^2_{{n,m}}}},\\
u^4_{{n,m},x,y}=2{{\rm e}^{-u^1_{{n,m}}+u^4_{{n,m}}}}-2{{\rm e}^{u^3_{{n,m}}-u^4_{{n,m}}}}
}
\eqs{\label{UXB21}
u^1_{{n+1,m},x}-u^1_{{n,m},x}=-2{{\rm e}^{u^4_{{n,m}}-u^1_{{n+1,m}}}}+2{{\rm e}^{u^1_{{n,m}}-u^2_{{n+1,m}}}},\\
u^2_{{n+1,m},x}-u^2_{{n,m},x}=-2{{\rm e}^{u^1_{{n,m}}-u^2_{{n+1,m}}}}+2{{\rm e}^{u^2_{{n,m}}-u^3_{{n+1,m}}}},\\
u^3_{{n+1,m},x}-u^3_{{n,m},x}=-2{{\rm e}^{u^2_{{n,m}}-u^3_{{n+1,m}}}}+2{{\rm e}^{u^3_{{n,m}}-u^4_{{n+1,m}}}},\\
u^4_{{n+1,m},x}-u^4_{{n,m},x}=-2{{\rm e}^{u^3_{{n,m}}-u^4_{{n+1,m}}}}+2{{\rm e}^{u^4_{{n,m}}-u^1_{{n+1,m}}}}
}
\eqs{\label{VYB21}
u^1_{{n,m+1},y}-u^1_{{n,m},y}=-2{{\rm e}^{u^1_{{n,m+1}}-u^2_{{n,m}}}}+2{{\rm e}^{u^4_{{n,m+1}}-u^1_{{n,m}}}},\\
u^2_{{n,m+1},y}-u^2_{{n,m},y}=-2{{\rm e}^{u^2_{{n,m+1}}-u^3_{{n,m}}}}+2{{\rm e}^{u^1_{{n,m+1}}-u^2_{{n,m}}}},\\
u^3_{{n,m+1},y}-u^3_{{n,m},y}=-2{{\rm e}^{u^3_{{n,m+1}}-u^4_{{n,m}}}}+2{{\rm e}^{u^2_{{n,m+1}}-u^3_{{n,m}}}},\\
u^4_{{n,m+1},y}-u^4_{{n,m},y}=-2{{\rm e}^{u^4_{{n,m+1}}-u^1_{{n,m}}}}+2{{\rm e}^{u^3_{{n,m+1}}-u^4_{{n,m}}}}}
\eqs{\label{VXB21}
u^1_{{n,m+1},x}-u^2_{{n,m},x}={{\rm e}^{u^1_{{n,m}}-u^1_{{n,m+1}}}}-{{\rm e}^{u^2_{{n,m}}-u^2_{{n,m+1}}}},\\
u^2_{{n,m+1},x}-u^3_{{n,m},x}={{\rm e}^{u^2_{{n,m}}-u^2_{{n,m+1}}}}-{{\rm e}^{u^3_{{n,m}}-u^3_{{n,m+1}}}},\\
u^3_{{n,m+1},x}-u^4_{{n,m},x}={{\rm e}^{u^3_{{n,m}}-u^3_{{n,m+1}}}}-{{\rm e}^{u^4_{{n,m}}-u^4_{{n,m+1}}}},\\
u^4_{{n,m+1},x}-u^1_{{n,m},x}={{\rm e}^{u^4_{{n,m}}-u^4_{{n,m+1}}}}-{{\rm e}^{u^1_{{n,m}}-u^1_{{n,m+1}}}}}
\eqs{\label{UYB21}
u^1_{{n+1,m},y}-u^4_{{n,m},y}={{\rm e}^{u^1_{{n+1,m}}-u^1_{{n,m}}}}-{{\rm e}^{u^4_{{n+1,m}}-u^4_{{n,m}}}},\\
u^2_{{n+1,m},y}-u^1_{{n,m},y}={{\rm e}^{u^2_{{n+1,m}}-u^2_{{n,m}}}}-{{\rm e}^{u^1_{{n+1,m}}-u^1_{{n,m}}}},\\
u^3_{{n+1,m},y}-u^2_{{n,m},y}={{\rm e}^{u^3_{{n+1,m}}-u^3_{{n,m}}}}-{{\rm e}^{u^2_{{n+1,m}}-u^2_{{n,m}}}},\\
u^4_{{n+1,m},y}-u^3_{{n,m},y}={{\rm e}^{u^4_{{n+1,m}}-u^4_{{n,m}}}}-{{\rm e}^{u^3_{{n+1,m}}-u^3_{{n,m}}}}}

\section*{Acknowledgments}

This work is partially supported by Russian Foundation for Basic Research (RFBR) grants $\#$ 11-01-97005-r-povoljie-a, $\#$ 12-01-92602-KO\_a and $\#$ 10-01-00088-a.


\begin{thebibliography}{EMG}

\bibitem{zamolodchikov} Belavin A. A., Polyakov A. M., Zamolodchikov A. B., Nucl. Phys., B241 (1984), 333–-373.

\bibitem{MOP} A. V. Mikhailov, M. A. Olshanetsky and A. M. Perelomov, Comm. Math. Phys. 79
(1981) 473.

\bibitem{OP} M. A. Olshanetsky, A. M. Perelomov, Physics Reports, Volume 71, Issue 5, May 1981, Pages 313-400.

\bibitem{M}  A. V. Mikhailov, Pis'ma Zh. Eksp. Teor.Fiz., 1979, V.30, N. 7, 443-448.

\bibitem{Bog} O.I.Bogoyavlensky, Commun. Math. Phys. 51 (1976) 201.

\bibitem{Cor} E.Corrigan, in: Particles and Fields, CRM Ser. Math. Phys., Vol. 1, Springer, New York, 1999.

\bibitem{LSSh1982} A.N. Leznov, V.G. Smirnov, A.B. Shabat, {\it Internal symmetry group and integrability conditions for two-dimensional dynamical systems}, Theoret. and Math. Phys. v. 51, no. 10, (1982).

\bibitem{DrS}
V. G. Drinfel'd, V. V. Sokolov, {\it Lie algebras and equations of Korteweg-de Vries type}, Journal of Mathematical Sciences V. 30, N. 2, 1975-2036

\bibitem{GanzhaTsarev} E.I. Ganzha, S.P. Tsarev, {\it Integration of classical series An, Bn, Cn, of exponential systems}, Krasnoyarsk, {2001}

\bibitem{ShabatYamilov}
A. B. Shabat, R. I. Yamilov, {\it Exponential systems of type I and
the Cartan matrices}, (In Russian), Preprint, Bashkirian Branch of Academy of Science of the USSR, Ufa, (1981).

\bibitem{Ward} R.S.Ward, {\it Discrete Toda field equations}, Phys. Lett. A (1995), 45-48.

\bibitem{Suris} Yu.B. Suris, {\it Generalized Toda chains in discrete time}, Leningrad Math. J., 2, {1990}, 339-352.


\bibitem{KNS}
A. Kuniba, T.Nakanishi, J.Suzuki, {\it $T$-systems and $Y$-systems in integrable systems}, J. Phys. A: Math. Theor. 44 (2011) 103001 (146pp)

\bibitem{IH}
R.Inoue, K.Hikami, {\it The lattice Toda field theory for simple Lie algebras: Hamiltonian structure and $\tau$-function}, Nuclear Physics B 581 [PM] (2000) 761–775

\bibitem{KNS2} A. Kuniba, T.Nakanishi, J.Suzuki: Int. J. Mod. Phys. A 9 (1994) 5215.

\bibitem{Tsuboi} Z. Tsuboi, {\it Solutions of Discretized Affine Toda Field Equations for $A^{(1)}_n$, $B^{(1)}_n$, $C^{(1)}_n$, $D^{(1)}_n$, $D^{(2)}_{n+1}$}, Journal of the Physical Society of Japan, v.66, No 11, 1997, pp. 3391-98.

\bibitem{AdlerStartsev} V. E. Adler, S. Ya. Startsev,
{\it On discrete analogues of the Liouville equation,} Teoret.
Mat. Fizika, \textbf{121}, no. 2, 271-284 (1999), (English
translation: Theoret. and Math. Phys. , \textbf{121}, no. 2,
1484-1495, (1999)).

\bibitem{H} I.T. Habibullin, {\it Discrete chains of the series C}, Theoret. and Math. Phys. 146 , no. 2, (2006) 170–182.


\bibitem{HZY}Ismagil Habibullin, Kostyantyn Zheltukhin, Marina Yangubaeva, {\it Cartan matrices and integrable lattice Toda field equations},//
Journal of Phys.A, \textbf{44} (2011).

\bibitem{GHY} Rustem Garifullin, Ismagil Habibullin, Marina Yangubaeva,  {\it Affine and finite Lie algebras and integrable Toda field equations on discrete space-time}, arXiv:1109.1689v3.
\bibitem{smirnov} Sergey V.Smirnov, {\it Semidiscrete Toda lattices}, 2012,	arXiv:1203.1764v1.

\bibitem{Zabr} Zabrodin A.V., {\it Hirota's difference equations}, Theoret. and Math. Phys. 113, no 2, {1997} 179-230.

\end{thebibliography}
\end{document}